\documentclass{emulateapj}
\usepackage{apjfonts}
\usepackage{lscape}
\newcommand{\rotate}{}

\shortauthors{White, Becker, Helfand}
\shorttitle{New Galactic Radio Catalogs}

\slugcomment{To appear in the August 2005 Astronomical Journal}

\begin{document}
\title{New Catalogs of Compact Radio Sources in the Galactic Plane}

\author{Richard~L.~White}
\affil{Space Telescope Science Institute, Baltimore, MD 21218}
\email{rlw@stsci.edu}

\author{Robert~H.~Becker}
\affil{Physics Dept., University of California, Davis, CA 95616\\
and IGPP/Lawrence Livermore National Laboratory
}
\email{bob@igpp.ucllnl.org}

\and 

\author{David J. Helfand}
\affil{Dept. of Astronomy, Columbia University, New York, NY 10027}
\email{djh@astro.columbia.edu}

\begin{abstract}
Archival data have been combined with recent observations of the
Galactic plane using the Very Large Array to create new catalogs
of compact centimetric radio sources.  The 20~cm source catalog
covers a longitude range of $-20^{\circ}<l<120^{\circ}$; the latitude
coverage varies from $\pm 0\fdg8$ to $\pm 2\fdg7$. The total survey
area is $\sim 331$~deg$^{2}$; coverage is 90\% complete at a flux
density threshold of $\sim 14$~mJy, and over 5000 sources are
recorded. The 6~cm catalog covers 43 deg$^{2}$ in the region
$-10^{\circ}<l<42^{\circ}, |b|<0\fdg4$ to a 90\% completeness
threshold of 2.9~mJy; over 2700 sources are found. Both surveys
have an angular resolution of $\sim6\arcsec$. These catalogs provide
a 30\% (at 20~cm) to 50\% (at 6~cm) increase in the number of
high-reliability compact sources in the Galactic plane, as well as
providing greatly improved astrometry, uniformity, and reliability;
they should prove useful for comparison with new mid- and far-infrared
surveys of the Milky Way.

\end{abstract}


\section{Introduction}

Between 1982 and 1991, the Very Large Array (VLA)\footnote{The
National Radio Astronomy Observatory is operated by Associated
Universities, Inc. under cooperative agreement with the National
Science Foundation} was used to conduct extensive snapshot surveys
along the plane of the Milky Way at both 6 and 20~cm. In a series
of papers in the early 1990s (Becker et al.\ 1990; Zoonematkermani
et al.\ 1990; White et al.\ 1991; Helfand et al.\ 1992; Becker et
al.\ 1992; Becker et al.\ 1994), we presented catalogs of over 4000
compact ($\theta\leq 25\arcsec$) radio sources, and used the best
far-infrared observations then available (IRAS) to construct large
samples of compact and ultracompact \ion{H}{2} regions, planetary nebule,
etc. Although we employed the best data reduction practices available
and tractable on the computers of that era, the images themselves
and the source catalogs derived from them left much to be desired.
Nonetheless, these data remain the most sensitive radio survey in
existence for compact radio emitters in the Galactic plane. The
{\sl Midcourse Space Experiment} (MSX)
mid-infared survey of the plane (Price et
al.\ 2001; Egan, Price, and Kramer 2003) offers significant
improvement in both sensitivity and angular resolution over the
largely source-confused IRAS images, and the recent launch of the
Spitzer Space Telescope presages extensive new mid- and far-IR
observations of the Milky Way.
These developments warrant a new
look at the available radio data.

Thus, we have carried out a complete re-reduction of the archival
VLA data, supplementing the $\sim3000$ individual pointings
with 28 hours of new observations designed to correct deficiencies
in the existing database. This paper presents new, improved images
and catalogs of discrete radio sources at both 6 and 20~cm, as well
as unveiling a new Web site that makes all the images publicly
accessible.

We have also undertaken a new multi-configuration 20~cm VLA map of
the Milky Way (D.~J.~Helfand et al., 2005, in preparation).  The new images
are of much higher quality than any previous radio observations of the
Galactic plane, but they cover only a portion of the plane
($5^{\circ}<l<32^{\circ}, |b|<0.6^{\circ}$) at a single
wavelength.  The older VLA data complement the new survey by
extending its area and frequency coverage.  While the analysis
of the new data is not yet complete, we use it in this paper in
several checks of the quality of the images and catalogs presented
here.

In section 2 we provide a description of the observations included
in this project, while \S 3 describes our data analysis, highlighting
the differences between the original reductions and our current
efforts. The source catalogs, containing over 10,000 entries,
comprise \S4, where we also provide descriptions of a number of
tests we have performed to quantify the surveys' astrometric and
photometric accuracies, as well as highlighting the caveats essential
for making productive use of these data. \S5 introduces the Galactic
plane Web site (\url{http://third.ucllnl.org/gps}), which allows
easy access to all images and catalogs, and describes some of the
uses to which these data products can (and will) be put.

\section{Observations}

\begin{figure*}
\plottwo{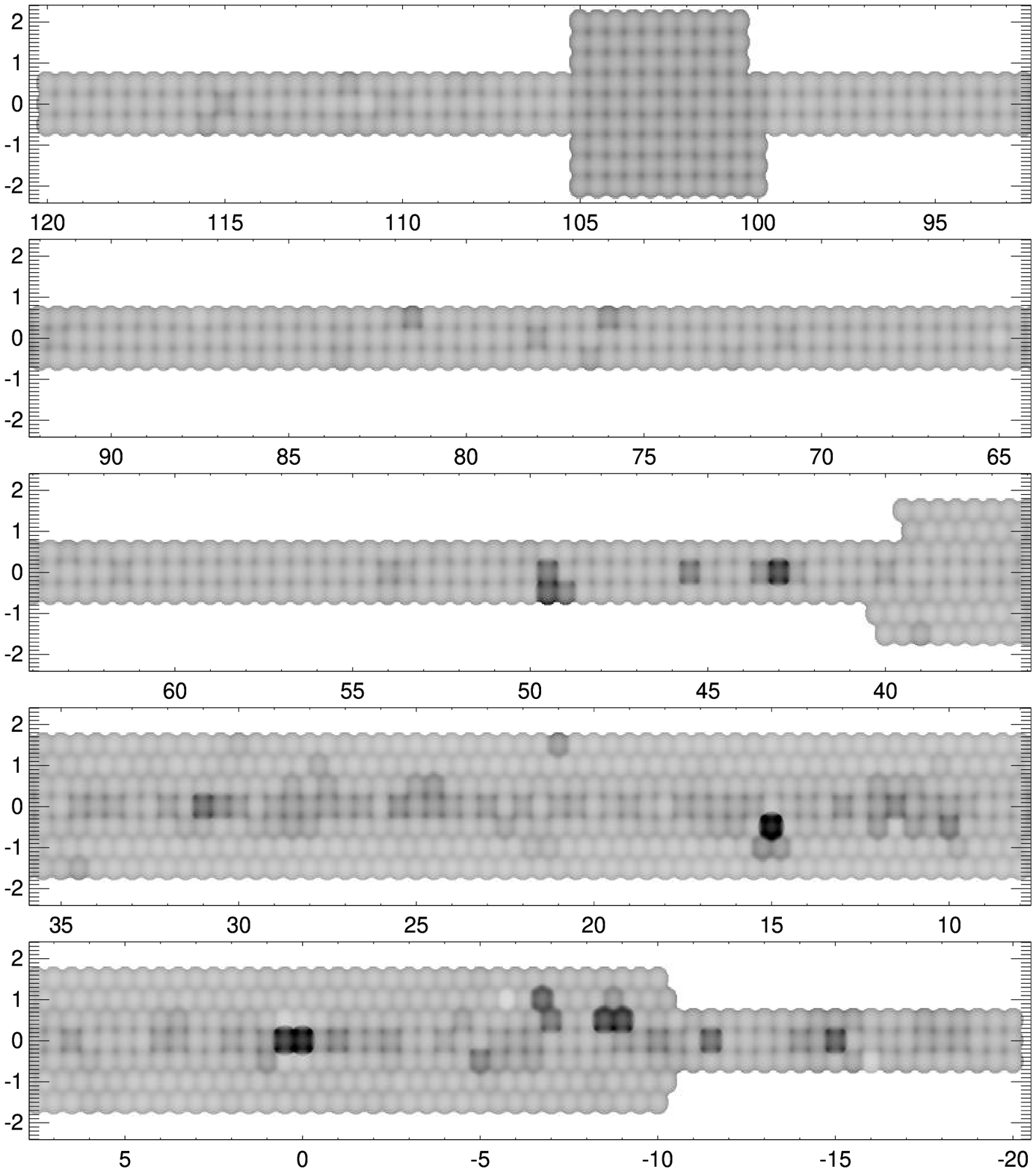}{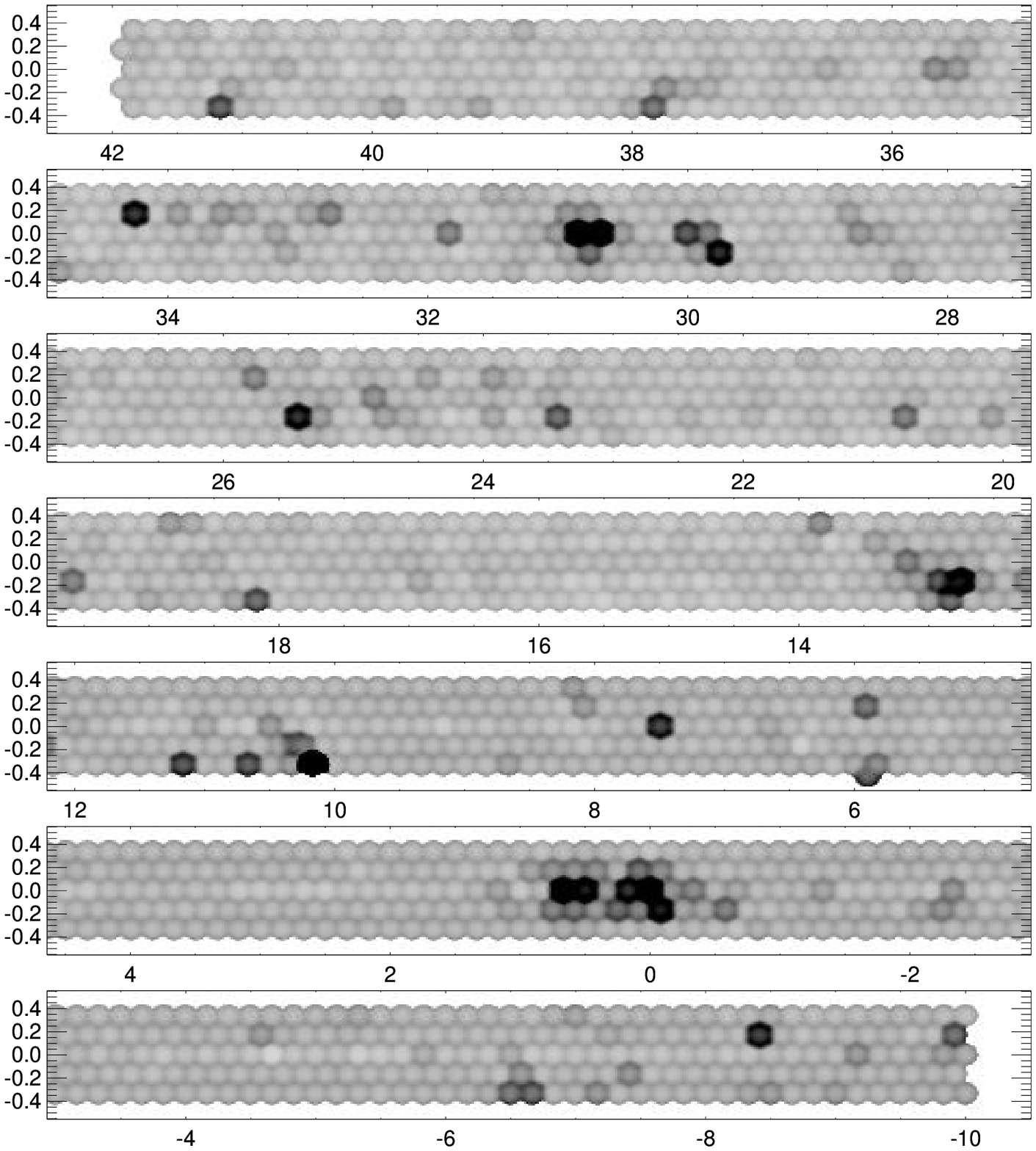}
\caption{
Coverage maps at 20~cm (left) and 6~cm (right).  The spatial scales
differ for the two panels.  Dark regions indicate areas of reduced
sensitivity due to elevated noise from bright sources.  The pattern
of pointing field centers is evident from the lower sensitivity
achieved far from the field centers.  Note that the oldest 20~cm
data (the three inner strips centered on the Galactic plane) did
not use staggered pointing positions on alternate rows, which leads
to greater variation in the sensitivity.
}
\label{fig-cover}
\end{figure*}

The observations used to construct the catalogs presented herein
were taken for a variety of purposes over a period of twenty-two
years in seven of the eight possible VLA configurations. The 20~cm
data acquired by other authors between 1982 and 1986 were on a
rectilinear grid that provided less-than-optimal coverage (see
Fig.~\ref{fig-cover}). Beginning in 1989, we sought, through a
series of proposals, to fill out the 20~cm coverage in the Galactic
center region and the first two quadrants, and to complement these
data in the inner Galaxy with a 6~cm snapshot survey. Some of the
latter data were corrupted when taken in the original C-configuration
and the observations were repeated in 1991, albeit in the
lower-resolution DnC and D configurations. Furthermore, some of the maps
constructed from the salvagable 1990 data were compromised by very
bright, extended sources. Given the continuing utility of our
compact source catalogs, we reobserved both the
low-resolution and compromised fields in the Spring 2004 C-configuration;
these data replace the respective 1990--1991 data in the current
analysis.  We also observed a single field at 20~cm in February 2005
to fill in the only remaining hole in the 20~cm coverage.

All of our analysis reported in this paper focuses on the total
intensity maps; we have not attempted to analyze the available
polarization data.  The 20~cm observations were mainly taken in
line mode to avoid the bandwidth smearing problems that accompany
wide-field continuum imaging at that wavelength.  Such data do not
have associated polarization information, so the available 20~cm
polarization data have incomplete sky coverage.  The 6~cm observations
were taken in continuum mode and could be used to construct
polarization images, but they would likely be of limited value
because most of the 6~cm Galactic emission is expected to be thermal
and unpolarized.  Another obstacle is that the data do not include
observations of polarization standard calibration sources.  Nonetheless,
it might be interesting for a future study to use these same data
to examine the polarization properties of Galactic radio sources.

The 20~cm observations taken in 1983 used one IF bandpass at 1611~MHz
that included a strong OH maser line.  Becker, White \& Proctor
(1992) used these data to construct a catalog of OH masers.  For
this paper we are interested in radio continuum emission, but it
is necessary in the data analysis to account for the OH masers.
The self-calibration particularly is complicated by the possibility
that a source may have very different flux densities in the two IFs
(see \S\ref{subsection-selfcal} below for further discussion.)

A summary of the observations used in our final analysis
is presented in Table~1.
We include the observation dates, configuration, number
of fields observed, Galactic longitude range covered, number of
bands and bandwidths used, and the central observing frequencies.
Most snapshots were of approximately 90-s duration. Over 110 hours
of observing time are represented by these observing programs (including
fields not used in our final images.)
The coverage is complete at 20~cm over the range
$-20^{\circ}<l<120^{\circ}$; the latitude coverage ranges from $\pm
0\fdg8$ to $\pm 2\fdg7$. The 6~cm coverage is complete in the region
$-10^{\circ}<l<42^{\circ}, |b|\lesssim0.4^{\circ}$. Total sky
coverage is approximately 331~deg$^2$ and 43~deg$^2$ at 20~cm and
6~cm, respectively. Figure~\ref{fig-cover} displays coverage maps
in a grayscale format that illustrates the different extents,
pointing grids, and sensitivities of the two surveys.

\begin{deluxetable}{ccrclc}
\tablecolumns{6}
\tablewidth{0pt}
\tabletypesize{\small}
\tablecaption{Observation log of 6~cm and 20~cm snapshot data\label{table-obs}}
\tablehead{
\colhead{Date} & \colhead{Config} & \colhead{\# fields} & \colhead{$l$ range} &
\colhead{channels${\times}\Delta\nu$} & \colhead{freq} \\
               &                  &                    & \colhead{(deg)}     &
\colhead{(MHz)} & \colhead{(MHz)}
}
\startdata
\cutinhead{20 cm Observations}\\
20 Jul 1982 &      B  &      96 &        100--105  &    1 $\times$ 6  &          1443 \\
24 Dec 1983 &    BnA  &      31 &        340--0    &    2 $\times$ 3  &        1441/1611 \\
27 Dec 1983 &    BnA  &      79 &        340--0    &    2 $\times$ 3  &        1441/1611 \\
30 Dec 1983 &    BnA  &     241 &        340--0    &    2 $\times$ 3  &        1441/1611 \\
24 Jul 1986 &      B  &     194 &        0--100    &    2 $\times$ 3  &        1441/1641 \\
25 Jul 1986 &      B  &     193 &        0--100    &    2 $\times$ 3  &        1441/1641 \\
23 Mar 1989 &      B  &      29 &        340--50   &    2 $\times$ 3  &        1465/1515 \\
27 Mar 1989 &      B  &      42 &        105--120  &    2 $\times$ 3  &        1465/1515 \\
29 Apr 1989 &      B  &       1 &        357.5     &    2 $\times$ 3  &          1465 \\
01 May 1989 &      B  &     199 &        350--40   &   14 $\times$ 3  &          1465 \\
02 May 1989 &      B  &     203 &        350--40   &   14 $\times$ 3  &          1465 \\
01 Feb 2005 &    BnA  &       1 &        354.25    &    2 $\times$ 25 &        1451/1490 \\
\cline{3-3}
Total       &         &    1309 &       & & \\ 
\cutinhead{6 cm Observations}\\
22 Jun 1989 &       C   &   228 &        350--18   &    2 $\times$ 50 &        4840/4890 \\
26 Sep 1990 &      CnB  &   139 &        350--13 & & \\
02 Oct 1990 &      CnB  &   136 &        350--13 & & \\
04 Oct 1990 &      CnB  &   138 &        350--13 & & \\
15 Oct 1990 &      CnB  &   145 &        350--13 & & \\
08 Dec 1990 &       C   &   255 &         14--42 & & \\
09 Dec 1990 &       C   &   257 &         13--42 & & \\
28 Feb 2004 &      CnB  &    61 &         26--40 & & \\
17 Apr 2004 &       C   &   113 &         22--42 & & \\
28 Apr 2004 &       C   &    91 &         10--42 & & \\
\cline{3-3}
Total       &           &  1563 & & & \\
\enddata
\tablecomments{Only observations used in final map construction are listed; bad data and superseded
observations are omitted.} 
\end{deluxetable}

\section{Data analysis}

Advances in both algorithms and computing power, along with the
new observations at 6~cm, allow us to produce significantly improved
images and catalogs from these Galactic plane survey data. All of
the editing, calibration, and subsequent analysis of these data
were accomplished using AIPS scripts of our own design employing
standard AIPS algorithms. Most of the improvements were taken from
the data processing pipeline developed for the FIRST survey (Becker,
Helfand \& White 1995; White et al.\ 1997).  We outline below the
significant differences in data processing between the original
analysis and the results derived here.

\subsection{Self-Calibration}
\label{subsection-selfcal}

No self-calibration was applied in the original analysis. Here, we
have utilized several iterations of self-calibration for all fields
containing a source `bright enough' to yield significant improvement
as defined by the criteria in the AIPS routine MAPIT; more than
half of all fields are now self-calibrated. This results in
significantly improved dynamic range over a majority of the survey
area.

Self-calibration is applied separately to the two frequency channels
of 20~cm fields that include OH masers from the catalog of
Becker, White \& Proctor (1992).  This is important because
the OH maser sources have very different brightnesses in
the two bandpasses, leading to errors if the channels are
self-calibrated jointly.  After self-calibration, the
channels are combined and mapped as a single image.

\subsection{Astrometric Distortion Correction}
\label{subsection-distortion}

No corrections were made in the original analysis for the image
distortions introduced by approximating the three-dimensional sky
as a two-dimensional plane. As noted in Helfand et al.\ (1992) this
produces map source positions offset from true positions by
$\sim3\arcsec$ at $20\arcmin$ from the pointing center, and up to
five times this value $28\arcmin$ off axis; correction factors are
calculable from the formula presented in Perley (1989) but were
not included in our published catalogs.  In the current analysis,
the AIPS task OHGEO was applied to all images, completely removing
these offsets and providing much improved astrometric accuracy.

\subsection{Image Co-adding}

In the original reduction, we did not take full advantage of the
significant overlap in coverage between adjacent images. Here the
images have been co-added to maximize sensitivity and minimize
variation in the survey sensitivity threshold (see Becker, Helfand
\& White 1995 for a complete discussion of the algorithm employed).
Since the observing grid was not optimized for co-adding, however,
the rms still rises by a factor of $\sim3$ at the boundaries between
fields (see Fig.~\ref{fig-cover} for details).

\subsection{Destriping}

Bright, extended radio sources are severely undersampled by
high-resolution, snapshot observations of the type reported here.
The common result is large-scale stripes through maps of regions
containing, or adjacent to, bright radio sources. In our original
analysis, no attempt was made to account for this nonuniformity in
the images. In the current work, we have applied a wavelet algorithm
to the images, using a high-pass filter to remove the worst of the
striping. The ``a trous'' wavelet transform (Starck \& Murtagh
1994) is used to decompose each field into a stack of images with
structures having increasing scale sizes.  The sharpest features
(point sources) are in the first level of the stack, features larger
by a factor of 2 in the second level, and so on.  The decomposition
is iterated to remove objects appearing in the sharper channels
from the lower-resolution levels of the stack, so the last level
of the stack contains an estimate of the large-scale `sky' (in this
case stripes from the deconvolution) underlying the sources.
Subtracting the smoothest channel from the data removes structures
larger than 1.5~arcmin from the image, which eliminates practically
no real features but does a good job in removing the stripes.

\subsection{Source Extraction}

In our original analysis, the source catalogs were constructed by
examining each of the $\sim3000$ images by eye and fitting a
two-dimensional Gaussian to each source appearing above the local
noise level.  This labor-intensive technique has the advantage of
allowing for an assessment of the reality of each potential source
(substantially reducing the number of sidelobes and other map
artifacts included as cataloged sources), but suffers from operator
error and subjectivity. Having spent considerable effort on the
development of an automated source detection algorithm for the
FIRST survey (dubbed HAPPY --- see White et al.\ 1997 for details),
we have applied this algorithm to our newly reduced images in order
to generate the catalogs reported here. Owing to the highly variable
noise levels in the plane and our nonuniform coverage, however, we
have modified our strategy in running HAPPY: we use a fainter search
threshold near field centers where the noise is lower. We accomplished
this by running our standard HAPPY algorithm on the same image
several times using different-sized windows of included area and
different flux density thresholds.  This allows us to detect sources
down to a $5\sigma$ flux density threshold everywhere in the maps.

\subsection{Sidelobe Probabilities}

The large amount of resolved flux, the high surface density of
bright extended sources in the plane, and the modest {\it u-v} coverage
inherent in snapshot observations lead to a significant sidelobe
contamination problem. As noted above, our original survey analysis
attempted to address this problem by examining each potential source
by eye and deciding whether it was a sidelobe. While this
approach was reasonably successful, it was far from foolproof\footnote{Or
Becker-proof.}; furthermore, the outcome of manual examination is
binary --- either a source is ignored as a sidelobe or is included
as a catalog source.

In the current analysis we have utilized an oblique decision tree
artificial intelligence algorithm (Murthy, Kasif \& Salzberg 1994)
to calculate for each source the probability that it is a sidelobe.
We used as a training set for this algorithm a deep, multi-configuration
map of the Galactic plane we are constructing at 20~cm (D.~J.~Helfand et
al., 2005, in preparation; see \S\ref{section-website}).
To date, this dataset covers the
region $5^{\circ}<l<32^{\circ}, |b|<0\fdg8$ (32 square degrees or $\sim
10\%$ of the existing 20~cm dataset) and encompasses 669 of the
20~cm sources detected by HAPPY in the current survey. The angular
resolutions of the two surveys are similar. The location of each
source in the current catalog was examined by eye in the new, deeper
images; missing sources were included as sidelobes in the
training set (except for those identified with OH masers
in the Becker, White \& Proctor [1992] catalog; see \S\ref{section-results}).
Since the threshold of the current catalog is $\sim
10$~mJy and that of the new survey images ranges from 1 to 2~mJy
(furthermore, the new images have much higher dynamic range and
greatly reduced sidelobe levels), there were essentially no ambiguous
cases; either a catalogued source was clearly present in the deeper
images or the field was blank. While it is conceivable that source
variability could explain the presence of a source in the current
catalog that was not apparent in the new maps, changes by factors
of five or more in flux density at 20~cm are extremely rare in both
extragalactic and Galactic radio sources; for statistical purposes,
it is safe to ignore this possibility.

\begin{figure*}
\epsscale{0.7}
\plotone{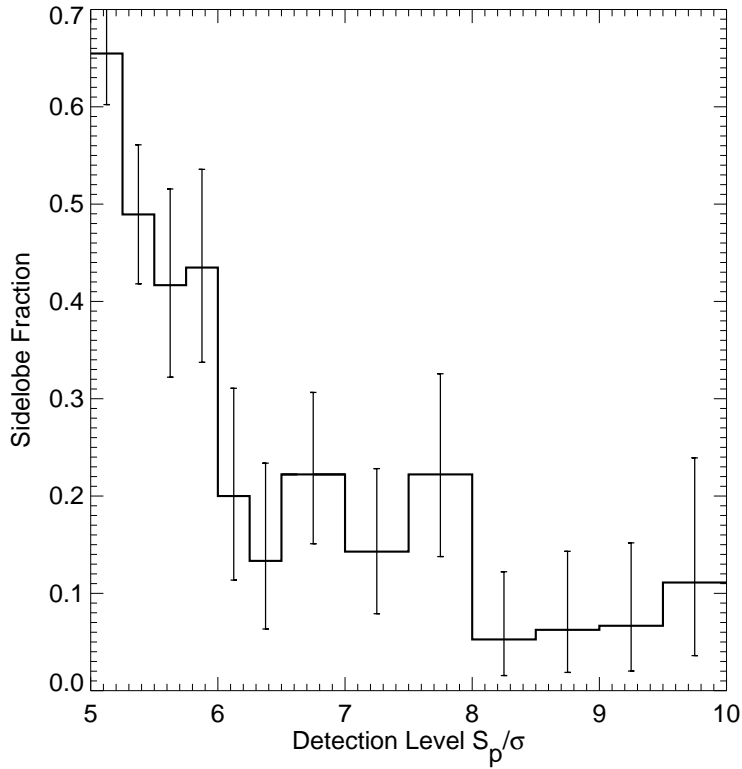}
\caption{
Sidelobe fraction versus detection level $S_p/$rms for the training set.
The fraction is 60\% for sources below $5.5\sigma$ and declines to
only 0.6\% for sources brighter than $10\sigma$.
}
\label{fig-sidelobefrac}
\end{figure*}

Of the 669 sources, 122 were identified as sidelobes.  The sidelobe
fraction is a strong function of source significance: of the sources
between 5 and $5.5\sigma$, nearly two-thirds are likely sidelobes,
whereas above $6\sigma$, this fraction falls to $<20\%$ (see
Fig.~\ref{fig-sidelobefrac}). Thus, we present below two catalogs
for each dataset: a primary catalog with a threshold of $5.5\sigma$,
and a `faint-source catalog' that extends the catalog to the
$5\sigma$ threshold.  We selected $5.5\sigma$ as the dividing line
since it is at this threshold that most of the sources added to
the catalog are real (see Fig.~\ref{fig-sidelobefrac}). The $5\sigma$
catalog contains many sources (perhaps a majority) that are sidelobes;
however, it also includes hundreds of real sources, and in applications
such as those that use a match to a catalog at another wavelength
as a filter for selecting real sources, the faint-source catalog
offers a valuable resource (e.g., see Giveon et al.\ 2005, which
presents a match to the mid-IR MSX catalog, and finds hundreds of
new compact and ultracompact \ion{H}{2} regions).

Only sources above $5.5\sigma$ are assigned a sidelobe probability.
We used the 538 training set sources above $5.5\sigma$ (including
44 sidelobes) to create decision tree classifiers using the OC1
oblique decision tree program (Murthy, Kasif \& Salzberg 1994).
The 15 parameters used for the classification include source
properties (peak to integrated flux ratios, rms noise levels, source
major and minor axes compared with the beam, etc.) and properties
of the nearest bright source that could be creating sidelobes
(positional offsets, flux ratios, etc.) Ten independent decision
trees were trained using the randomized search method, and the
weighted classifications from those trees were combined to obtain
a sidelobe probability estimate using the same approach adopted by
White et al.\ (2000).  The accuracy of the classification and the
probability estimates were confirmed using 5-fold cross validation
(also described in White et al.)

The sidelobe probabilities $P_s$ indicate that 79\% of the 20~cm
sources are highly reliable ($P_s<0.1$).  An additional 9\% are
fairly reliable ($0.1<P_s<0.25$), 6.0\% are unreliable ($0.25<P_s<0.5$),
and 6.3\% are most probably sidelobes ($P_s>0.5$).  Overall we
estimate that about 10\% of the $5.5\sigma$ 20~cm catalog sources
are sidelobes.

At 6~cm, no comparable `truth set' exists, so it is not possible
to train our decision tree algorithm to recognize sidelobes. The
notion of using the new 20~cm images for comparison was explored,
but given the possibility (indeed, the certainty) that inverted
spectrum ultracompact \ion{H}{2} regions will appear in the 6~cm maps
and be absent at 20~cm, we decided to apply the algorithm developed
for the 20~cm data with the simple scalings expected for the
differences between the 6 and 20~cm images (mainly due to the
$5\times$ lower rms in the 6~cm images, which allows considerably
fainter sources to create sidelobes.) We do not have high confidence
in this approach, and we advise caution when interpreting the 6~cm
sidelobe probabilities. Again, we present two catalogs, one with
a threshold of $5.5\sigma$, and a faint-source compendium reaching
to the $5\sigma$ threshold.

\subsection{Summary}

The net result of all these improvements is an increase in the
number of detected $5.5\sigma$ sources from 3406 to 5084 at 20~cm
(4006 with low sidelobe probabilities, $P_s<0.1$, of which 28\%
are newly detected sources) and from 1272 to 2729 at 6~cm (1986
with low sidelobe probabilities, of which 47\% are new), along with
significant improvements in coverage uniformity, survey sensitivity,
and astrometric accuracy.  The comparison between the old and new
catalogs is summarized in Table~2.

\begin{deluxetable*}{ccccccccccccc}
\tablecolumns{10}
\tabletypesize{\scriptsize}
\tablewidth{0pt}
\tabcolsep=3pt
\rotate
\tablecaption{Comparison of Old and New Catalogs}
\tablehead{
\colhead{Band} &
\colhead{Survey} &
\colhead{Reference} &
\colhead{$\ell, b$} &
\multicolumn{3}{c}{Original} &
\multicolumn{3}{c}{New}\\
($\lambda$) & & & (deg) & \# scrs & threshold & (compl.) & \# scrs & threshold & (compl.)\\
& & & & & (mJy) & (\%) & & (mJy) & (\%)}
\startdata   
20~cm &    I & Zoonematkermani & $-20<\ell <120, |b|<0.8$ &     1992 &   8-25 & 75\%\\
&    & et al. (1990) & & & & \\
      &   II & Helfand et al. (1992)  &    $-10<\ell<40, 0.8<|b|<1.8$ &  1457 &  5--20 & 95\%\\
& total & & & $\overline{3406}$\tablenotemark{a} & & &  $\overline{6919}$  &   13.8 & 90\%\\
& & & & & & &($\sim5000$)\tablenotemark{b}\\
6~cm &  III &  Becker et al. (1994) &        $-10<\ell<40, |b|<0.4$ &      1272 & 2.5--10 & 98\%  & 3283  &    2.9  & 90\%\\
&        +IV & this work  &          $-10<\ell<40, |b|<0.4$ (590 fields) & & & &             ($\sim 2500$)\tablenotemark{b}\\
\enddata   
\tablenotetext{a}{As noted in Helfand et al. (1992), 43 sources were common between the two catalogs; the reported total takes this into account.}
\tablenotetext{b}{Approximate number of real sources, excluding sidelobes.}
\end{deluxetable*}

\section{The Survey Results}
\label{section-results}

\subsection{Four Catalogs}

In Tables 3--6 we present our new catalogs of compact radio sources
in the Galactic plane. Table~3 is a complete list of the 5084 20~cm
sources detected at a significance of $>5.5\sigma$. We include the
Galactic longitude and latitude, RA (J2000), Dec (J2000), peak and
integrated flux densities, the computed rms in the map at the source
position, major and minor axes and position angles derived from
elliptical Gaussian fits, the name of the field containing the
source, and the probability that the source is a sidelobe. The
final column contains a flag `o' indicating whether or not the source
was in our original catalog (Zoonematkermani et al.\ 1990; Helfand
et al.\ 1992). In addition, 19 sources in Table~3 and one in Table~4
have `OH' flags indicating
that they are not continuum sources but instead are 
detected through their 1611~MHz OH maser emission
(Becker, White \& Proctor 1992).
In Table~4, we include the same information (excluding
the sidelobe probability) for the 1835 sources falling between 5.0
and $5.5\sigma$. Again, we emphasize that roughly 60\% of these
sources are likely to be sidelobes; this faint-source catalog should only be
used in conjunction with other catalogs that help filter out the true
sources.

The 6~cm survey covers only $\sim 13\%$ of the 20~cm survey area
but is about five times deeper than the 20~cm survey. All 20~cm
sources with spectral indices less steep than $\alpha=-1.35$ (where
$S_{\nu}\propto \nu^{\alpha}$) --- the vast majority of all Galactic
and extragalactic objects --- should have a 6~cm counterpart. Thus,
in presenting the catalogs for the 6~cm survey, we include columns
recording the 20~cm peak and integrated flux densities, the computed
map rms, and the sidelobe probability.  Table~5 presents the 2729
sources detected at greater than $5.5\sigma$ significance. It also
includes the 179 20~cm sources that fall in the 6~cm survey area
but lack 6~cm counterparts (discussed further below).  Three of the
sources in Table~5 are fainter than $5.5\sigma$ at 6~cm but have
20~cm counterparts, which in our judgment makes them reliable, and
we include them here.  And five 6~cm sources in the table are listed
twice because they match two 20~cm counterparts within 10~arcsec.
In total the table has 2916 entries.

The columns in Table~5 are similar to those in Tables~3--4 with
the addition of the 20~cm data.  The rms flux density is listed
for both surveys even when a source is not present at both wavelengths.
The position and field name come from the 6~cm survey except for
20~cm-only sources.  The flag indicating whether a source was in
the original catalog (last column) here has four possible values:
`c' (in the old 6~cm catalog), `l' (in the old 20~cm catalog), `cl'
(in both catalogs), and blank (in neither catalog).  The `OH' flag is
repeated here for masers (none of which are detected at 6~cm.)
The flag also has an asterisk for a few 20~cm-only sources very close to the
6~cm survey edge (see below.)

Table~6 lists the 551 5.0 to $5.5\sigma$ 6~cm sources whose
reliability is less certain.  Its columns are identical to those
in Table~4.  None of these sources have 20~cm counterparts since
the few low-reliability sources with matches are included in Table~5.

\subsection{Quality Assessment}

We have examined by eye the 114 20~cm sources lacking 6~cm counterparts
that fall within the region covered by our new, deep, multi-array
survey. Of these, 55 (48\%) are in regions of diffuse emission where
the source detection algorithm has chosen a different component
structure at the two frequencies; in such cases, nearby catalog
entries should be considered as a single source complex. Of these
sources, 12 fall in the $S_p<5.5\sigma$ sample, while 41 of the
remaining 43 (95\%) have, appropriately, low sidelobe probabilities
($<0.25$). Seven of the sources are detected in the 20~cm catalog
because they are OH masers, which are not expected to have 6~cm
continuum emission.  A total of 44 (39\%) of the 20~cm sources
lacking 6~cm counterparts are almost certainly spurious (sidelobes,
noise bumps barely above threshold, etc.); of these, 41/44 (93\%)
have sidelobe probabilities above 0.25 or are found in the
low-reliability ($S_p<5.5\sigma$) sample.  Thus, it appears from
this comparison datset that using 0.25 as the sidelobe probability
threshold is both $>90$\% accurate and $>90$\% complete. The remaining
eight 20~cm sources without 6~cm matches do appear in our deep,
multi-array 20~cm images. In a few cases, they fall very close to
the boundary of the 6~cm coverage where the sensitivity is too low
for them to be detected, but the majority of cases are truly
steep-spectrum and/or highly variable sources.  The former objects
are noted with an asterisk in Table 5.

A detailed source-by-source comparison between the old and new 6~cm
catalogs was carried out for the region $5^{\circ}<l<15^{\circ}$
using both the 20 and 6~cm images, the new multi-array 20~cm survey
images, and the MSX images.  In this $\sim 8$~deg$^2$ region, 198
6~cm sources appear in the old catalog; 171 (86\%) of these are
also found in the new catalog, albeit with slightly shifted (and
improved) positions and revised flux densities. Of the 27 missing
sources, 15 are spurious objects that have disappeared in the
reprocessed images and three are real sources that now simply fall
outside the trimmed survey area. Examination of the new 20~cm data
shows that five missing sources are actually peaks in a large
diffuse region of thermal emission evident in the $20~\mu$m MSX
map; these are no longer considered significant in the reprocessed
high-resolution (and rather noisy) 6~cm image, and so do not appear
in the new catalog. The final four missing sources are almost
certainly real and are present in the images, but fall below the
formal threshold for acceptance to the catalog.

This same region contains 23 sources from the old 20~cm catalog
that had no match in the old 6~cm catalog. Eight are still
found in the new 20~cm catalog, with five of those now having
counterparts in the new 6~cm catalog.  The three without new 6~cm
counterparts include an OH maser, a source that falls in the
low-sensitivity region at the edge of the 6~cm survey area, and a
truly steep-spectrum (or variable) source, 9.001+0.078, that falls
just below the threshold for catalog inclusion but is clearly present
in the reprocessed 6~cm image.  Of the 15 old 20~cm sources that
are absent from the new 20~cm catalog, 11 are clearly spurious since
they fail completely to appear in our deep multi-array images.  The
other four lie in diffuse source complexes that are largely resolved
out at 6~cm, although one of those actually does appear in the new
6~cm catalog.

The new 6~cm catalog contains 489 entries with greater than
$5.5\sigma$ significance in the $5^\circ < l < 15^\circ$ region,
more than double the number of sources in the old catalog.
While a number of these new entries represent components of large
source complexes previously grouped as one source, there are also
many new sources added as a consequence of the new data and improved
processing techniques. In summary, then, our revised analysis has
removed 26 spurious sources from the old catalog while losing only
seven real discrete objects (which can be recovered from examination
of the images), and has more than doubled the number of sources
and source components detected.

\subsection{Survey Sensitivity}

\begin{figure*}
\epsscale{1.0}
\plottwo{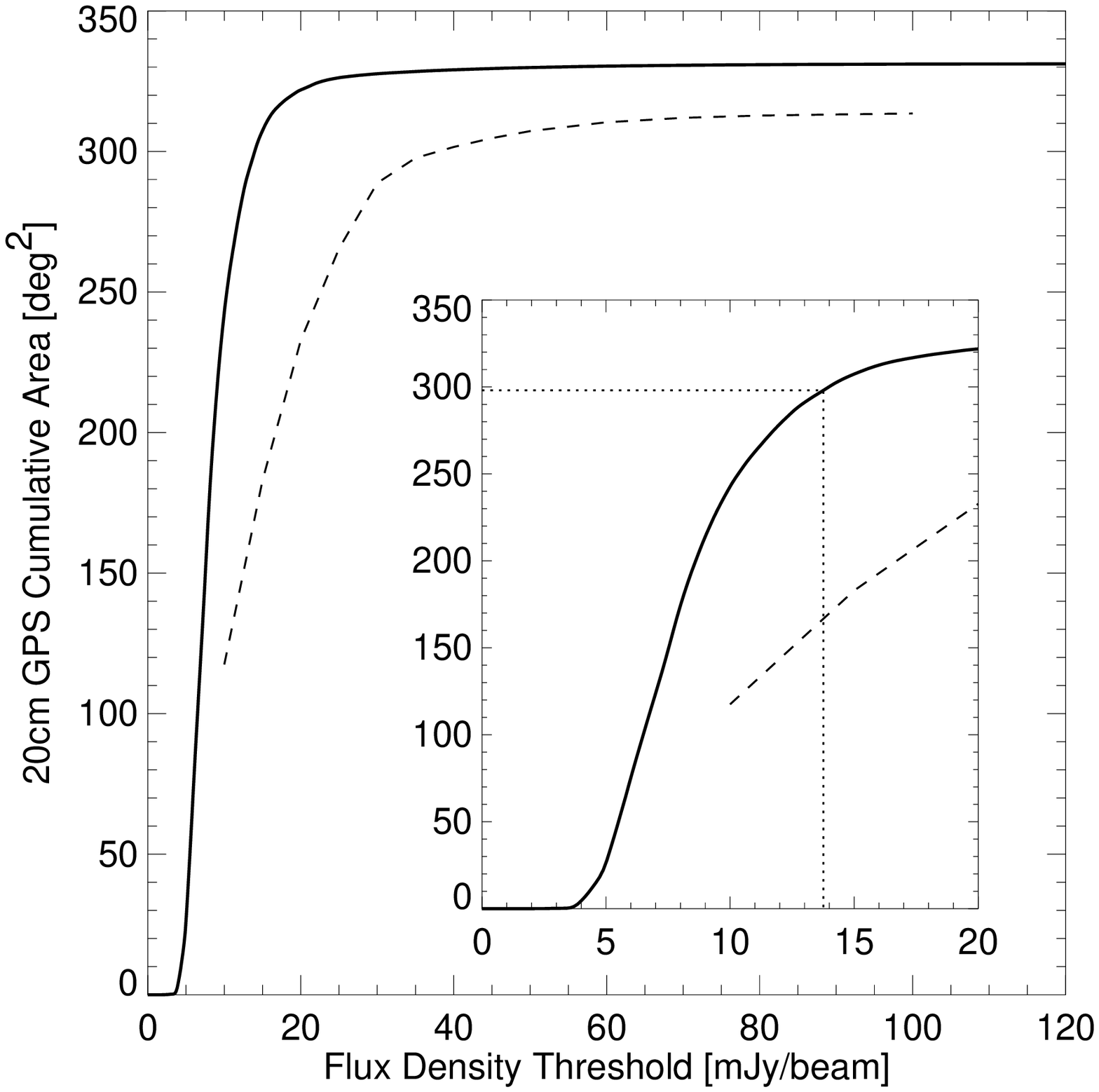}{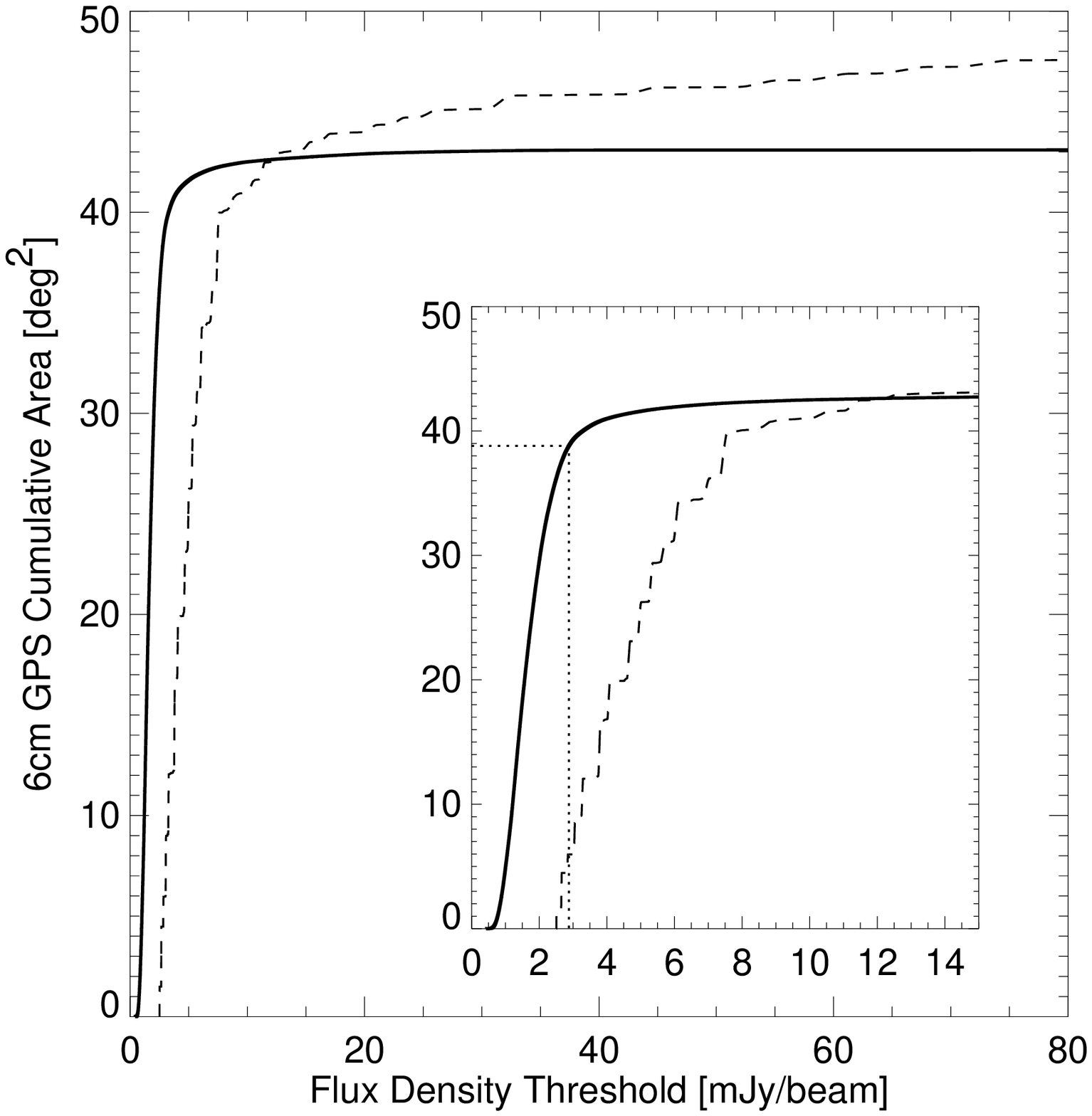}
\caption{
Cumulative area functions for the 20~cm (left) and 6~cm surveys.  The $y$-axis
is the sky area covered to the given sensitivity limit or better at a $5.5\sigma$
detection threshold.  The dashed lines show the coverage from the old versions of
these surveys, while the dotted lines indicate the 90\% completeness levels.
}
\label{fig-cumarea}
\end{figure*}

Due to the variety of observing modes used and the sparse observing
grid, the sensitivity over the survey area varies significantly.
The median rms values are 0.897~mJy and 0.179~mJy for the 20 and
6~cm surveys, respectively; with our improved reduction techniques
and addition of new data, only $\sim 5\%$ of the fields at both
wavelengths exceed these median values by more than a factor of
two (42 out of 1309 fields at 20~cm and 90 out of 1563 fields at
6~cm). In Figure~\ref{fig-cumarea} we show the cumulative coverage
for the two surveys as a function of flux density threshold for
the high-reliability catalogs. A useful figure of merit to describe
the surveys is that 90\% of the survey areas have thresholds deeper
than 2.9~mJy and 13.8~mJy for 6 and 20~cm, respectively.

\begin{figure*}
\epsscale{0.7}
\plotone{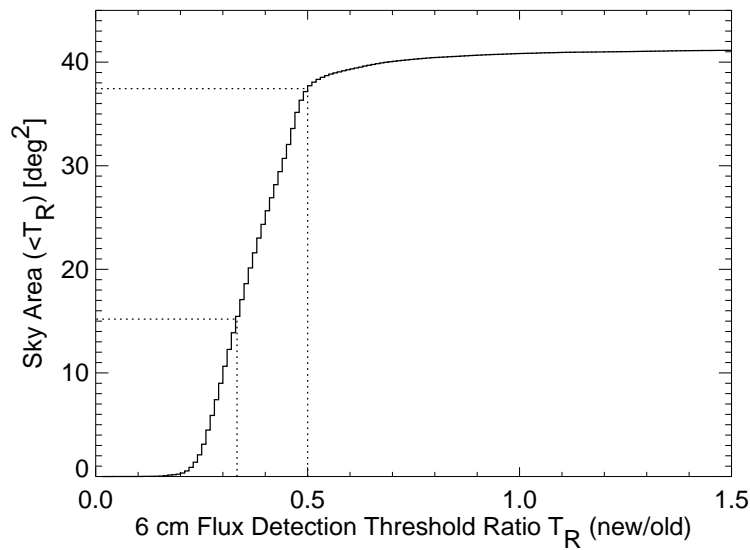}
\caption{
Improvement in the flux density detection limits for the 6~cm survey.
The dotted lines mark factors of 2 and 3 improvement in sensitivity.
}
\label{fig-detectionthreshold}
\end{figure*}

Also shown in Figure~\ref{fig-cumarea} is the marked improvement
in sensitivity that has resulted from our new analysis. The dashed
curves represent the cumulative coverage of the original surveys.
Two points are worth noting. First, for the 6~cm survey, it is
clear that the total effective area in the old analysis exceeds
that in the new at high flux density thresholds ($>$10--20~mJy).
This is a consequence of the fact that we have trimmed the edges
of the survey region, dropping all areas where the sensitivity
falls to less than $\sim 10\%$ of the on-axis value; a total of 56
sources appearing in the old catalog --- all with poorly determined
positions and flux densities because they are far from a pointing
center --- are dropped in the new catalog. A few of these sources
actually appear in our new images, but are not in the catalog as
a consequence of the fact that the HAPPY source detection algorithm
excludes sources if their source extraction island intersects the
boundary of the image. More importantly, the lower flux density
thresholds of the current analysis are apparent in the figure. In
Figure~\ref{fig-detectionthreshold}, we quantify the change for
the 6~cm survey by plotting the cumulative sky area as a function
of the flux detection threshold ratio of the new maps compared to
the old. Fully 40\% of the survey area shows a factor of three
improvement, while 90\% gains at least a factor of two.

\subsection{Survey Astrometry}

\begin{figure*}
\epsscale{0.7}
\plotone{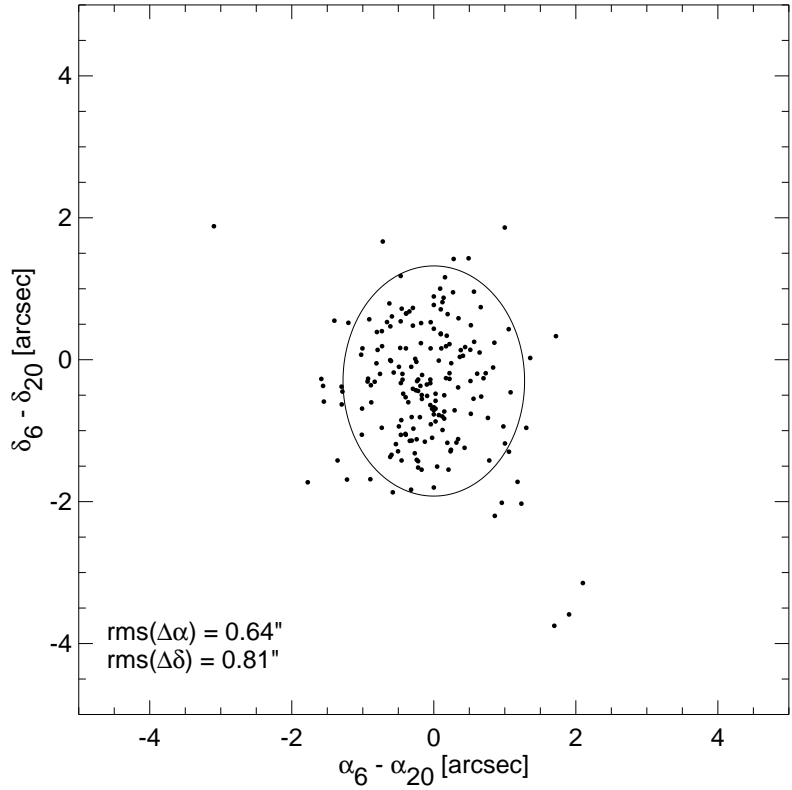}
\caption{
Position differences for sources appearing in both the 6~cm and 20~cm surveys.
Only sources that are compact with deconvolved major axes smaller
than 5~arcsec are included.  The $2\sigma$ error ellipse is shown.
The scatter is larger in declination due to the extension of the
synthesized VLA beam in the north-south direction at southern
declinations.
}
\label{fig-deltapos}
\end{figure*}

As noted in \S\ref{subsection-distortion}, a major improvement in the new catalogs is the
inclusion of a correction for the distortion introduced into VLA
images by mapping the three-dimensional sky onto a two-dimensional
image; some sources in the original catalogs had astrometric errors
exceeding $5\arcsec$ as a consequence of this effect. A simple
assessment of the astrometric accuracy of the current catalogs is
provided by examining the offsets between the point sources (major
axes $<5\arcsec$) found in both catalogs. We display this comparison
in Figure~\ref{fig-deltapos}.  The rms position discrepancies are
$0.64\arcsec$ in RA and $0.81\arcsec$ in declination. Since
uncertainties in both positions are included in this comparison,
we infer that the rms position errors for the individual catalogs
are $\Delta \alpha = 0.45\arcsec$
and $\Delta\delta = 0.57\arcsec$. Even these values should be
treated as upper limits, since spectral index variability over even
compact sources can induce centroid shifts that masquerade as astrometric
errors.

A further test of the astrometry comes from comparing our positions
to the deep multi-configuration dataset described above.  There
are 212 sources at 20~cm and 300 sources at 6~cm that match nearly
point-like sources in the new catalog.  The rms positional errors
for the 20~cm catalog are $\Delta\alpha = 0.67\arcsec$,
$\Delta\delta=0.78\arcsec$, and for the 6~cm catalog are $\Delta\alpha
= 0.67\arcsec$, $\Delta\delta=0.84\arcsec$, These are consistent
with the values quoted above given that the deeper survey also has
positional errors.  The median position offsets between the two
surveys are $\Delta\alpha = -0.05\arcsec$, $\Delta\delta=0.10\arcsec$
(20~cm) and $\Delta\alpha = -0.04\arcsec$, $\Delta\delta=-0.14\arcsec$
(6~cm).

\begin{figure*}
\epsscale{0.7}
\plotone{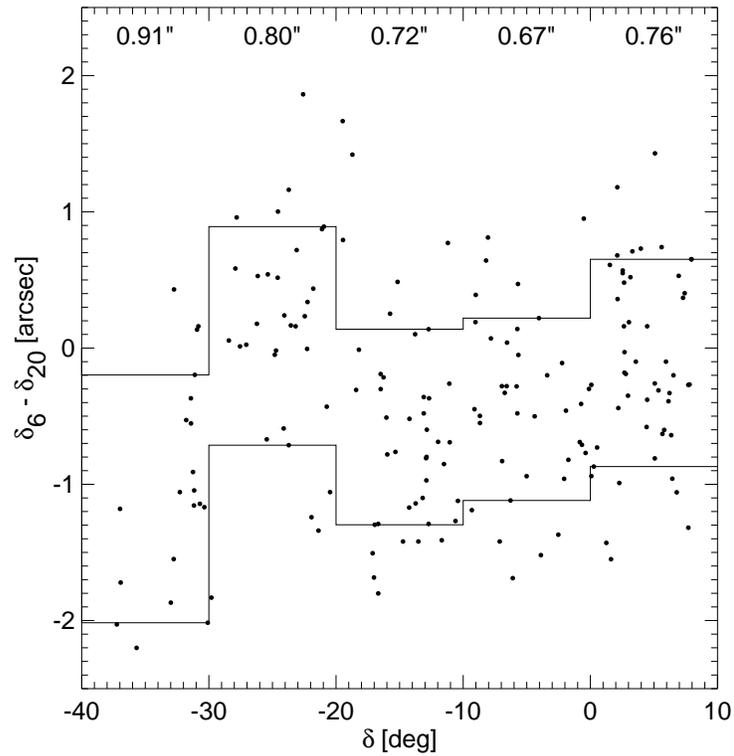}
\caption{
Difference between 6~cm and 20~cm declinations as a function of
source declination.  The lines indicate the $\pm1\sigma$ range about
the mean, and the numbers at the top give the value of $\sigma$ for each
bin.  The increase in the uncertainty for more
southerly sources is apparent, as is a small systematic offset 
of $\sim 0.25\arcsec$ between the two catalogs.
}
\label{fig-posdec}
\end{figure*}

The greater positional uncertainty in declination is occasioned by
the fact that the VLA synthesized beam is elongated in the north-south
direction when observing sources at southerly declinations; this
effect is apparent in Figure~\ref{fig-posdec} where it is seen that
the scatter in declination increases by $\sim 30\%$ going from
sources with $\delta>-10^{\circ}$ to the southern limit of the
survey at $\delta \sim -37^{\circ}$.  Also apparent in these figures is
a $\sim 0.25\arcsec$ shift to the south for the 20~cm catalog with
respect to the 6~cm catalog; the comparison to our new deep survey
shows that the offsets for the two surveys are both approximately
0.1\arcsec\ but in opposite directions.  This sets the absolute
astrometric accuracy of the catalogs presented here.  The origin
of these small shifts is unclear, and further tests are under way to
understand it.

\subsection{Survey Photometry}

\begin{figure*}
\epsscale{0.7}
\plotone{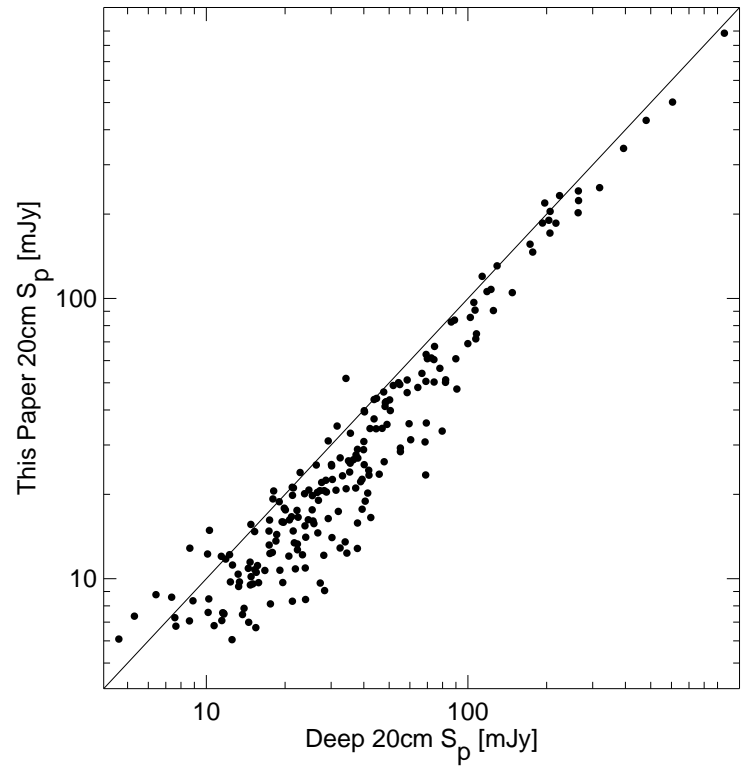}
\caption{
Comparison of 20~cm flux densities from this paper to a deep
multi-configuration catalog for 212 compact sources.  The
agreement is good.  The distribution is skewed toward
fainter flux densities in this paper's catalog because the
multi-configuration images recover more flux from slightly
extended sources and do not suffer from ``CLEAN bias''.
}
\label{fig-fluxes}
\end{figure*}

We expect the photometry in these catalogs to be superior to the
original analysis because the noise in the maps is reduced.  We
have checked the photometric accuracy of the 20~cm catalog using
point-source counterparts in the deep multi-configuration catalog
(see above).  The results are shown in Figure~\ref{fig-fluxes}.
The general agreement between flux densities in the two surveys is
good.  The scatter is clearly asymmetric, with flux densities from
this paper's catalog tending to fall below those from the
multi-configuration catalog.  This can be attributed to the tendency
of our single-configuration snapshot observations to resolve out
flux from slightly extended sources.  The ``CLEAN bias'' effect
also reduces the flux densities of fainter sources in snapshot
surveys (White et al.\ 1997).  We have not attempted to correct
for the CLEAN bias (as we did for the FIRST survey) since we lack the
data to model the effect accurately in these complex, highly variable
maps.  But users of the catalog should be cognizant of the likely
underestimate of flux densities in the catalog due to the bias.

\goodbreak
\section{The Galactic Plane Website}
\label{section-website}
\nobreak

While catalogs are a convenient and compact form in which to present
the primary results of these surveys, the images themselves are
also of great utility.  In addition to their use in making overlays
with observations at different wavelengths and in assessing the
validity of a given catalog entry, the vast majority of the more
than two billion pixels comprising the images are noise --- but not
noise without content. Each of the 235 million beam areas that does
not contain a source provides an upper limit to the radio flux
density for any object at that location. Furthermore, stacking this
`noise' at the locations of many sources identified in another
wavelength band can provide the mean radio flux density for the
source class in question to levels far below the typical image rms
(e.g., Glikman et al.\ 2004).

Consistent with our past practice of providing user-friendly access
to source images and catalogs (e.g., our VLA FIRST survey), we are
making available all of these images on the MAGPIS Web site
(\url{http://third.ucllnl.org/gps}).  The {\bf M}ulti-{\bf A}rray
{\bf G}alactic {\bf P}lane {\bf I}maging {\bf S}urvey collects, as
its namesake is wont to do, bits and pieces of the Galactic sky
that have been imaged at high resolution. At its inception, the
site includes the 6 and 20~cm data described here, as well as the
main MAGPIS database, which currently includes high-dynamic range,
high-sensitivity images for the region $5^{\circ}<l<32^{\circ},
|b|<0.6^{\circ}$ (D.~J.~Helfand et al., 2005, in preparation). Much of this
latter area is being imaged with XMM-Newton at hard-X-ray wavelengths,
and all of it has been mapped at mid-infrared wavelengths by MSX;
mosaics of the latter data, gridded onto the same coordinate system
as the radio images, are included at this site, as will be the
X-ray data as they become available.  The high-resolution,
high-sensitivity GLIMPSE Legacy Project, currently being conducted
with the Spitzer Space Telescope, includes much of the same region
of sky, and will undoubtedly increase significantly the value of
these radio images.

Even though this work comprises a Galactic survey, there is a
substantial admixture of extragalactic sources. At a threshold
peak flux density of $S_{20cm} = 13.8$~mJy, our 90\% completeness
threshold, the FIRST survey (Becker,
Helfand \& White 1995) --- a high-latitude extragalactic survey with
similar angular resolution --- predicts 9.0 extragalactic radio
sources per square degree compared with the 10.3 sources deg$^{-2}$
with $S_p>13.8$~mJy in the 20~cm catalog presented here.  Even in the Galactic
plane, then, 88\% of the compact radio sources are extragalactic.
In some cases extragalactic sources can be classified as such by
their morphology (e.g., radio doubles), but for the most part
selecting out the Galactic sources requires multiwavelength followup
observations. The combination of these VLA data with the near-IR
(2MASS), mid- and far-IR (MSX and Spitzer) and X-ray (XMM-Newton)
databases now being assembled promises a substantially improved
view of the source populations and activity in Galactic regions
obscured from view at visible wavelengths.

\acknowledgments

RHB and DJH acknowledge the support of the National Science Foundation
under grants AST-02-6-309 and AST-02-6-55, respectively. RHB's work
was supported in part under the auspices of the US Department of
Energy by Lawrence Livermore National Laboratory under contract
W-7405-ENG-48.  DJH was also supported in this work by NASA grant
NAG5-13062. RLW acknowledges the support of the Space Telescope
Science Institute, which is operated by the Association of Universities
for Research in Astronomy under NASA contract NAS5-26555.

\clearpage

\begin{landscape}
\begin{deluxetable*}{rrrrrrrrrrrc}
\tablecolumns{12}
\tablewidth{0pc}
\rotate
\tabletypesize{\scriptsize}
\tablecaption{20~cm Sources With $S_p>5.5\sigma$}
\tablehead{
\colhead{Name ($l+b$)} & \colhead{RA} & \colhead{Dec} & 
\colhead{$P_s$\tablenotemark{a}} & \colhead{$S_p$} & \colhead{$S_i$} & \colhead{$\sigma_S$} & 
\colhead{Maj} & \colhead{Min} & \colhead{PA} & \colhead{Field} & 
\colhead{Notes\tablenotemark{b}} \\
\colhead{$^\circ$} & \colhead{(2000)} & \colhead{(2000)} & 
\colhead{} & \colhead{mJy} & \colhead{mJy} & \colhead{mJy} & 
\colhead{\arcsec} & \colhead{\arcsec} & \colhead{$^\circ$} & \colhead{Name} & 
\colhead{} \\
\colhead{(1)} & \colhead{(2)} & \colhead{(3)} & 
\colhead{(4)} & \colhead{(5)} & \colhead{(6)} & \colhead{(7)} & 
\colhead{(8)} & \colhead{(9)} & \colhead{(10)} & \colhead{(11)} & 
\colhead{(12)}
}
\startdata
 26.280${-}$0.932 & 18 42 35.605 & ${-}$06 20 39.03 & 0.41 &      4.59 &       5.61 &    0.729 &    3.79 &    1.25 &   21.9 & 262${-}$10 &      \\
 26.317+0.410 & 18 37 51.951 & ${-}$05 41 48.92 & 0.02 &     26.14 &      30.34 &    1.408 &    2.51 &    0.00 &  108.8 & 265+05 & o    \\
 26.318+0.412 & 18 37 51.677 & ${-}$05 41 39.67 & 0.02 &     14.02 &      15.35 &    1.394 &    2.62 &    0.00 &   74.0 & 265+05 & o    \\
 26.318${-}$1.673 & 18 45 19.304 & ${-}$06 38 53.45 & 0.54 &      8.01 &       4.61 &    1.361 &    1.11 &    0.00 &   82.6 & 265${-}$15 &      \\
 26.327${-}$1.531 & 18 44 49.720 & ${-}$06 34 33.36 & 0.02 &    162.39 &     173.04 &    0.891 &    1.87 &    0.00 &   51.6 & 265${-}$15 & o    \\
 26.345+1.316 & 18 34 40.922 & ${-}$05 15 19.14 & 0.02 &     19.07 &      27.68 &    1.264 &    4.13 &    2.04 &   59.8 & 265+15 & o    \\
 26.367${-}$0.905 & 18 42 39.619 & ${-}$06 15 16.35 & 0.02 &     23.49 &      30.41 &    0.895 &    5.02 &    0.00 &   30.6 & 265${-}$05 & o    \\
 26.367${-}$0.902 & 18 42 38.840 & ${-}$06 15 09.85 & 0.02 &      8.40 &       7.20 &    0.903 &    2.50 &    0.00 &   73.9 & 265${-}$05 & o    \\
 26.377+1.730 & 18 33 15.934 & ${-}$05 02 11.38 & 0.03 &     10.65 &      11.76 &    1.580 &    3.51 &    0.00 &   68.2 & 265+15 &      \\
 26.381+1.678 & 18 33 27.499 & ${-}$05 03 23.00 & 0.02 &     12.88 &       8.59 &    1.165 &    0.00 &    0.00 &   78.1 & 260+15 & o    \\
 26.382+0.971 & 18 35 59.048 & ${-}$05 22 51.66 & 0.02 &      6.48 &       6.67 &    0.859 &    2.82 &    0.00 &   72.0 & 267+10 & o    \\
 26.398${-}$0.498 & 18 41 15.736 & ${-}$06 02 25.96 & 0.02 &     13.30 &      13.65 &    0.977 &    2.07 &    0.00 &  158.7 & 265${-}$05 & o    \\
 26.428${-}$0.044 & 18 39 41.614 & ${-}$05 48 20.76 & 0.02 &     15.55 &      15.57 &    1.770 &    1.28 &    0.00 &   34.4 & 265+00 & o,OH \\
 26.430${-}$1.684 & 18 45 33.791 & ${-}$06 33 14.26 & 0.02 &     17.84 &      17.39 &    1.038 &    2.47 &    0.00 &   76.7 & 260${-}$15 & o    \\
 26.436+0.059 & 18 39 20.351 & ${-}$05 45 06.18 & 0.02 &     75.42 &     102.64 &    1.786 &    4.50 &    2.18 &  170.8 & 265+00 & o    \\
 26.436+0.826 & 18 36 36.063 & ${-}$05 23 58.21 & 0.02 &     24.40 &      27.67 &    1.442 &    3.06 &    0.00 &   70.1 & 262+10 & o    \\
 26.448+1.743 & 18 33 21.142 & ${-}$04 58 02.78 & 0.02 &     11.90 &      57.90 &    1.441 &   14.63 &    6.59 &  133.7 & 265+15 & o    \\
 26.450${-}$1.285 & 18 44 10.280 & ${-}$06 21 14.55 & 0.02 &      7.32 &       7.66 &    1.136 &    4.98 &    0.00 &   62.7 & 267${-}$10 &      \\
 26.451${-}$0.937 & 18 42 55.763 & ${-}$06 11 39.68 & 0.69 &      6.12 &       3.61 &    1.065 &    0.77 &    0.00 &   86.3 & 267${-}$10 &      \\
 26.452+0.560 & 18 37 34.666 & ${-}$05 30 28.48 & 0.78 &      6.68 &      66.05 &    0.970 &   18.89 &   12.57 &  106.9 & 265+05 &      \\
 26.460${-}$0.050 & 18 39 46.411 & ${-}$05 46 48.17 & 0.14 &      9.62 &      15.94 &    1.708 &    5.81 &    2.79 &  152.4 & 265+00 &      \\
 26.479+1.648 & 18 33 44.899 & ${-}$04 59 01.77 & 0.49 &      4.76 &      12.73 &    0.861 &    7.08 &    6.02 &  108.6 & 270+15 &      \\
 26.494+1.578 & 18 34 01.516 & ${-}$05 00 07.67 & 0.24 &      4.21 &       2.85 &    0.711 &    2.79 &    0.00 &   84.0 & 265+15 &      \\
 26.495${-}$1.749 & 18 45 54.984 & ${-}$06 31 29.97 & 0.36 &      8.77 &       6.40 &    1.415 &    2.07 &    0.00 &   73.4 & 265${-}$15 &      \\
 26.509${-}$0.567 & 18 41 42.623 & ${-}$05 58 22.27 & 0.62 &      5.05 &       3.97 &    0.908 &    3.67 &    0.00 &   97.6 & 265${-}$05 &      \\
\enddata
\tablecomments{Table 3 is published in its entirety in the electronic edition of the {\it Astronomical
Journal} and is also available on the MAGPIS website (\url{http://third.ucllnl.org/gps}).  A portion is
shown here for guidance regarding its form and content.}
\tablenotetext{a}{$P_s$ is the probability that the source is a sidelobe
of a nearby bright object (see the text for details.)}
\tablenotetext{b}{
Source notes: `o' indicates that the source was
in the original catalogs of Zoonematkermani et al.\ (1990) or
Helfand et al.\ (1992).  `OH' marks sources detected through 1611~MHz
OH maser emission rather than 20~cm continuum emission (Becker, White
\& Proctor 1992).
}
\end{deluxetable*}

\clearpage
\end{landscape}

\begin{landscape}
\begin{deluxetable*}{rrrrrrrrrrc}
\tablecolumns{11}
\tablewidth{0pc}
\rotate
\tabletypesize{\scriptsize}
\tablecaption{20~cm Sources With $S_p<5.5\sigma$}
\tablehead{
\colhead{Name ($l+b$)} & \colhead{RA} & \colhead{Dec} & 
\colhead{$S_p$} & \colhead{$S_i$} & \colhead{$\sigma_S$} & 
\colhead{Maj} & \colhead{Min} & \colhead{PA} & \colhead{Field} & 
\colhead{Notes\tablenotemark{a}} \\
\colhead{$^\circ$} & \colhead{(2000)} & \colhead{(2000)} & 
\colhead{mJy} & \colhead{mJy} & \colhead{mJy} & 
\colhead{\arcsec} & \colhead{\arcsec} & \colhead{$^\circ$} & \colhead{Name} & 
\colhead{} \\
\colhead{(1)} & \colhead{(2)} & \colhead{(3)} & 
\colhead{(4)} & \colhead{(5)} & \colhead{(6)} & 
\colhead{(7)} & \colhead{(8)} & \colhead{(9)} & \colhead{(10)} & 
\colhead{(11)}
}
\startdata
 24.819${-}$1.161 & 18 40 43.438 & ${-}$07 44 52.85 &      5.09 &       5.99 &    1.018 &    4.80 &    0.00 &   66.4 & 247${-}$10 &      \\
 24.835${-}$1.002 & 18 40 11.041 & ${-}$07 39 38.81 &      3.95 &       4.39 &    0.761 &    3.61 &    0.00 &   52.2 & 247${-}$10 &      \\
 24.839${-}$1.442 & 18 41 46.166 & ${-}$07 51 28.58 &      4.81 &       8.14 &    0.953 &    5.96 &    0.00 &   78.6 & 250${-}$15 &      \\
 24.846${-}$1.485 & 18 41 56.353 & ${-}$07 52 17.68 &      4.83 &       4.10 &    0.896 &    2.85 &    0.00 &   78.5 & 250${-}$15 &      \\
 24.847${-}$1.638 & 18 42 29.230 & ${-}$07 56 27.17 &      6.23 &       6.55 &    1.169 &    6.23 &    0.00 &   88.6 & 250${-}$15 &      \\
 24.849+0.087 & 18 36 18.318 & ${-}$07 08 54.78 &     12.44 &     154.02 &    2.331 &   23.36 &   13.50 &  153.6 & 250+05 &      \\
 24.874${-}$1.642 & 18 42 33.105 & ${-}$07 55 08.55 &      5.89 &       6.31 &    1.077 &    5.48 &    0.00 &   92.3 & 250${-}$15 &      \\
 24.881+1.056 & 18 32 53.935 & ${-}$06 40 26.35 &      4.94 &       3.89 &    0.907 &    2.06 &    0.00 &   81.2 & 250+15 &      \\
 24.887${-}$1.178 & 18 40 54.631 & ${-}$07 41 43.83 &      6.31 &       5.77 &    1.238 &    4.36 &    0.00 &   76.5 & 252${-}$10 &      \\
 24.902+1.230 & 18 32 19.037 & ${-}$06 34 31.23 &      6.77 &       4.39 &    1.353 &    1.68 &    0.00 &   78.4 & 247+10 &      \\
 24.907+1.057 & 18 32 56.697 & ${-}$06 39 00.51 &      5.18 &       5.98 &    0.967 &    4.13 &    0.00 &   92.3 & 250+15 &      \\
 24.938+0.883 & 18 33 37.508 & ${-}$06 42 11.41 &      6.36 &       4.74 &    1.217 &    2.72 &    0.00 &   86.0 & 247+10 &      \\
 24.968+1.642 & 18 30 58.162 & ${-}$06 19 35.47 &      4.20 &       5.91 &    0.818 &    6.93 &    0.00 &   87.7 & 250+15 &      \\
 25.026+1.562 & 18 31 21.837 & ${-}$06 18 43.05 &      3.62 &       3.56 &    0.670 &    5.61 &    0.00 &   81.7 & 250+15 &      \\
 25.046+0.673 & 18 34 34.486 & ${-}$06 42 15.75 &     12.41 &      33.33 &    2.409 &    8.22 &    4.43 &   77.5 & 252+10 & o    \\
 25.111${-}$1.027 & 18 40 47.004 & ${-}$07 25 35.95 &      4.85 &       6.71 &    0.942 &    5.09 &    0.00 &   67.1 & 252${-}$10 &      \\
 25.168${-}$0.923 & 18 40 31.009 & ${-}$07 19 43.40 &      4.42 &      13.63 &    0.878 &   15.14 &    0.00 &  128.5 & 247${-}$10 &      \\
 25.169+1.595 & 18 31 30.709 & ${-}$06 10 11.24 &      5.02 &       5.81 &    0.963 &    5.27 &    0.00 &   73.4 & 255+15 &      \\
 25.172${-}$1.213 & 18 41 33.625 & ${-}$07 27 28.03 &      7.08 &       5.94 &    1.328 &    2.64 &    0.00 &   56.3 & 252${-}$10 &      \\
 25.201${-}$0.945 & 18 40 39.303 & ${-}$07 18 34.62 &      4.41 &       5.96 &    0.808 &    6.73 &    0.00 &   98.9 & 252${-}$10 &      \\
 25.289+1.610 & 18 31 40.786 & ${-}$06 03 23.71 &      5.71 &       3.90 &    1.138 &    1.63 &    0.00 &   76.7 & 255+15 &      \\
 25.351${-}$1.409 & 18 42 35.714 & ${-}$07 23 17.40 &      4.56 &       4.18 &    0.908 &    3.73 &    0.00 &   65.9 & 255${-}$15 &      \\
 25.429+0.887 & 18 34 31.281 & ${-}$06 15 55.24 &      6.23 &       4.48 &    1.199 &    3.59 &    0.00 &   76.8 & 252+10 &      \\
 25.444${-}$0.853 & 18 40 46.378 & ${-}$07 03 06.05 &      7.41 &      14.57 &    1.423 &   10.71 &    0.00 &   75.8 & 255${-}$05 &      \\
 25.463${-}$1.248 & 18 42 13.407 & ${-}$07 12 52.65 &      6.88 &      25.44 &    1.366 &   17.57 &    1.36 &  139.2 & 252${-}$10 &      \\
\enddata
\tablecomments{Table 4 is published in its entirety in the electronic edition of the {\it Astronomical
Journal} and is also available on the MAGPIS website (\url{http://third.ucllnl.org/gps}).  A portion is
shown here for guidance regarding its form and content.}
\tablenotetext{a}{
Source notes are the same as for Table~3.
}
\end{deluxetable*}

\clearpage
\end{landscape}

\begin{landscape}
\begin{deluxetable*}{rrrrrrrrrrrrrrrrl}
\tablecolumns{17}
\tablewidth{0pc}
\rotate
\tabletypesize{\scriptsize}
\tablecaption{6~cm Sources With $S_p>5.5\sigma$}
\tablehead{
\colhead{} & \colhead{} & \colhead{} & 
\multicolumn{7}{c}{6 cm Data} & \colhead{} & \multicolumn{4}{c}{20 cm Data} \\
\cline{4-10} \cline{12-15} \\
\colhead{Name ($l+b$)} & \colhead{RA} & \colhead{Dec} & 
\colhead{$P_s$\tablenotemark{a}} & \colhead{$S_p$} & \colhead{$S_i$} & \colhead{$\sigma_S$} & 
\colhead{Maj} & \colhead{Min} & \colhead{PA} & 
\colhead{} & \colhead{$P_s$\tablenotemark{a}} & \colhead{$S_p$} & \colhead{$S_i$} & \colhead{$\sigma_S$} & 
\colhead{Field} & \colhead{Notes\tablenotemark{b}} \\
\colhead{$^\circ$} & \colhead{(2000)} & \colhead{(2000)} & 
\colhead{} & \colhead{mJy} & \colhead{mJy} & \colhead{mJy} & 
\colhead{\arcsec} & \colhead{\arcsec} & \colhead{$^\circ$} & 
\colhead{} & \colhead{} & \colhead{mJy} & \colhead{mJy} & \colhead{mJy} & 
\colhead{Name} & \colhead{} \\
\colhead{(1)} & \colhead{(2)} & \colhead{(3)} & 
\colhead{(4)} & \colhead{(5)} & \colhead{(6)} & \colhead{(7)} & 
\colhead{(8)} & \colhead{(9)} & \colhead{(10)} & 
\colhead{} & \colhead{(11)} & \colhead{(12)} & \colhead{(13)} & \colhead{(14)} & 
\colhead{(15)} & \colhead{(16)}
}
\startdata
354.829+0.077 & 17 32 25.653 & ${-}$33 16 08.52 & 0.02 &     10.32 &      16.24 &    0.365 &    6.18 &    3.24 &  170.4 &&&&&    1.464 & 35483+00 & c    \\
354.832${-}$0.409 & 17 34 23.112 & ${-}$33 31 53.60 &&&&    0.457 &&&  && 0.62 &     23.42 &      29.85 &    4.091 &  3550+00 &      \\
354.871${-}$0.006 & 17 32 52.111 & ${-}$33 16 48.21 & 0.02 &      4.25 &       5.72 &    0.217 &    4.37 &    0.00 &  119.1 &&&&&    1.136 & 35483+00 & c    \\
354.871${-}$0.012 & 17 32 53.513 & ${-}$33 16 56.58 & 0.02 &      1.78 &       3.69 &    0.221 &    8.62 &    4.01 &  156.8 &&&&&    1.130 & 35483+00 & c    \\
354.892+0.025 & 17 32 47.954 & ${-}$33 14 40.43 & 0.02 &      5.44 &      11.91 &    0.287 &    7.05 &    5.51 &  159.1 &&&&&    1.069 & 35483+00 & c    \\
354.934+0.329 & 17 31 41.623 & ${-}$33 02 36.55 &&&&    0.371 &&&  && 0.02 &      9.67 &      44.23 &    1.329 &  3550+05 &      \\
354.936+0.328 & 17 31 41.990 & ${-}$33 02 32.90 & 0.02 &      9.34 &      82.72 &    0.358 &   21.13 &   11.79 &   86.4 &&&&&    1.347 & 35500+33 &      \\
354.937+0.330 & 17 31 41.881 & ${-}$33 02 27.48 & 0.02 &      9.19 &     145.08 &    0.354 &   26.03 &   20.06 &  145.8 &&&&&    1.340 & 35483+33 &      \\
354.938+0.333 & 17 31 41.294 & ${-}$33 02 16.15 & 0.02 &     17.45 &      50.52 &    0.347 &   10.75 &    5.38 &   36.8 && 0.02 &     34.21 &     113.57 &    1.307 & 35500+33 & l    \\
354.939+0.332 & 17 31 41.734 & ${-}$33 02 15.71 & 0.02 &     40.99 &     101.75 &    0.344 &    8.45 &    3.26 &   94.0 &&&&&    1.325 & 35483+33 & c    \\
354.940+0.328 & 17 31 42.732 & ${-}$33 02 19.73 & 0.02 &     16.26 &      72.15 &    0.341 &   18.40 &    6.50 &  166.5 && 0.02 &     16.36 &      85.77 &    1.316 & 35483+33 &      \\
354.963+0.016 & 17 33 01.371 & ${-}$33 11 26.00 & 0.02 &      5.96 &       7.14 &    0.189 &    5.37 &    0.00 &  163.4 &&&&&    0.948 & 35500+00 & c    \\
354.973+0.416 & 17 31 26.867 & ${-}$32 57 47.34 &&&&    0.614 &&&  && 0.52 &      6.17 &       9.31 &    1.015 &  3550+05 &      \\
354.977+0.304 & 17 31 54.496 & ${-}$33 01 16.61 & 0.02 &     10.24 &      10.33 &    0.272 &    1.88 &    0.00 &  170.7 && 0.02 &     22.07 &      32.38 &    1.360 & 35500+33 & cl   \\
354.982${-}$0.209 & 17 33 58.287 & ${-}$33 17 47.79 & 0.02 &      4.84 &       5.49 &    0.427 &    4.00 &    0.00 &   40.9 &&&&&    2.412 & 35500${-}$33 & c    \\
355.000${-}$0.027 & 17 33 17.461 & ${-}$33 10 59.04 & 0.02 &      4.50 &       5.62 &    0.169 &    5.95 &    0.60 &  167.1 &&&&&    0.940 & 35500+00 & c    \\
355.008${-}$0.195 & 17 33 59.039 & ${-}$33 16 01.76 & 0.14 &      2.39 &       1.80 &    0.396 &    2.11 &    0.00 &   76.5 &&&&&    2.107 & 35500${-}$33 &      \\
355.068${-}$0.303 & 17 34 34.527 & ${-}$33 16 32.86 &&&&    0.417 &&&  && 0.18 &     24.92 &      27.75 &    3.985 &  3550+00 &      \\
355.105+0.097 & 17 33 04.043 & ${-}$33 01 37.75 & 0.02 &      9.92 &       9.15 &    0.371 &    0.51 &    0.00 &   91.7 && 0.02 &     31.30 &      29.87 &    1.195 & 35516+00 & cl   \\
355.110${-}$0.028 & 17 33 34.777 & ${-}$33 05 28.25 & 0.02 &      2.54 &       4.86 &    0.248 &    6.16 &    4.78 &   21.0 &&&&&    1.080 & 35516+00 &      \\
355.112${-}$0.023 & 17 33 33.832 & ${-}$33 05 11.48 & 0.52 &      1.43 &       4.82 &    0.234 &   13.37 &    6.38 &  175.6 &&&&&    1.083 & 35533+00 &      \\
355.127${-}$0.030 & 17 33 37.849 & ${-}$33 04 38.63 & 0.61 &      1.38 &       1.54 &    0.211 &    2.10 &    0.00 &   98.9 &&&&&    1.140 & 35516+00 &      \\
355.129${-}$0.303 & 17 34 43.966 & ${-}$33 13 26.00 & 0.02 &     10.68 &      12.45 &    0.300 &    2.61 &    0.85 &  120.8 &&&&&    4.819 & 35516${-}$33 & c    \\
355.198+0.130 & 17 33 10.484 & ${-}$32 55 51.35 & 0.02 &      4.68 &      19.32 &    0.395 &   15.63 &    6.39 &   31.6 &&&&&    1.797 & 35533+00 &      \\
355.203${-}$0.016 & 17 33 46.371 & ${-}$33 00 25.25 & 0.02 &      2.32 &       5.39 &    0.190 &    7.40 &    4.21 &   92.4 &&&&&    1.439 & 35516+00 & c    \\
\enddata
\tablecomments{Table 5 is published in its entirety in the electronic edition of the {\it Astronomical
Journal} and is also available on the MAGPIS website (\url{http://third.ucllnl.org/gps}).  A portion is
shown here for guidance regarding its form and content.}
\tablenotetext{a}{$P_s$ is the probability that the source is a sidelobe
of a nearby bright object (see the text for details.)}
\tablenotetext{b}{
Source notes: `c' indicates that the source was
in the original 6~cm catalog of Becker et al.\ (1994).
`l' indicates that the source was
in the original 20~cm catalogs of Zoonematkermani et al.\ (1990) or
Helfand et al.\ (1992).  `OH' marks sources detected through 1611~MHz
OH maser emission rather than 20~cm continuum emission (Becker, White
\& Proctor 1992).
}
\end{deluxetable*}

\clearpage
\end{landscape}

\begin{landscape}
\begin{deluxetable*}{rrrrrrrrrrc}
\tablecolumns{11}
\tablewidth{0pc}
\rotate
\tabletypesize{\scriptsize}
\tablecaption{6~cm Sources With $S_p<5.5\sigma$}
\tablehead{
\colhead{Name ($l+b$)} & \colhead{RA} & \colhead{Dec} & 
\colhead{$S_p$} & \colhead{$S_i$} & \colhead{$\sigma_S$} & 
\colhead{Maj} & \colhead{Min} & \colhead{PA} & \colhead{Field} & 
\colhead{Notes\tablenotemark{a}} \\
\colhead{$^\circ$} & \colhead{(2000)} & \colhead{(2000)} & 
\colhead{mJy} & \colhead{mJy} & \colhead{mJy} & 
\colhead{\arcsec} & \colhead{\arcsec} & \colhead{$^\circ$} & \colhead{Name} & 
\colhead{} \\
\colhead{(1)} & \colhead{(2)} & \colhead{(3)} & 
\colhead{(4)} & \colhead{(5)} & \colhead{(6)} & 
\colhead{(7)} & \colhead{(8)} & \colhead{(9)} & \colhead{(10)} & 
\colhead{(11)}
}
\startdata
 22.727+0.269 & 18 31 42.065 & ${-}$08 56 52.20 &      1.77 &       1.69 &    0.341 &    4.40 &    0.00 &   83.2 & 2283+33 &      \\
 22.749+0.308 & 18 31 36.236 & ${-}$08 54 37.57 &      1.51 &       1.79 &    0.291 &    5.84 &    0.00 &   75.3 & 2266+33 &      \\
 22.750${-}$0.248 & 18 33 36.141 & ${-}$09 09 57.64 &      1.37 &       4.11 &    0.273 &   11.88 &    5.85 &  165.3 & 2283+00 &      \\
 22.761${-}$0.239 & 18 33 35.448 & ${-}$09 09 09.21 &      1.22 &       1.54 &    0.232 &    4.81 &    0.00 &  105.1 & 2266${-}$33 &      \\
 23.186+0.162 & 18 32 56.515 & ${-}$08 35 25.47 &      1.14 &       2.22 &    0.224 &    9.39 &    0.00 &   70.9 & 2316+33 &      \\
 23.203+0.142 & 18 33 02.985 & ${-}$08 35 03.70 &      1.06 &       4.98 &    0.203 &   13.31 &    8.24 &   87.2 & 2333+33 &      \\
 23.224${-}$0.137 & 18 34 05.382 & ${-}$08 41 40.76 &      1.05 &       1.55 &    0.198 &    9.27 &    0.00 &  167.5 & 2316${-}$33 &      \\
 23.435${-}$0.204 & 18 34 43.347 & ${-}$08 32 18.40 &      5.84 &      18.11 &    1.092 &   13.43 &    5.34 &   16.9 & 2350${-}$33 & o    \\
 23.479+0.069 & 18 33 49.475 & ${-}$08 22 22.34 &      2.48 &       6.18 &    0.471 &    7.13 &    6.57 &   39.1 & 2350+00 &      \\
 23.621+0.374 & 18 32 59.781 & ${-}$08 06 22.03 &      0.82 &       0.84 &    0.164 &    2.69 &    0.00 &   74.1 & 2366+33 &      \\
 23.649${-}$0.039 & 18 34 31.787 & ${-}$08 16 18.52 &      0.98 &       3.61 &    0.191 &   20.62 &    3.62 &   18.4 & 2366+00 &      \\
 23.690+0.342 & 18 33 14.371 & ${-}$08 03 35.64 &      0.63 &       0.61 &    0.115 &    3.68 &    0.00 &   79.6 & 2366+33 &      \\
 23.747+0.103 & 18 34 12.141 & ${-}$08 07 09.18 &      1.87 &       2.27 &    0.353 &    7.91 &    0.00 &  168.0 & 2383+00 &      \\
 23.822+0.392 & 18 33 18.401 & ${-}$07 55 10.53 &      1.23 &       8.98 &    0.225 &   16.49 &   12.77 &  114.1 & 2383+33 &      \\
 24.017+0.238 & 18 34 13.101 & ${-}$07 49 05.03 &      2.05 &       8.01 &    0.391 &   14.69 &    6.43 &  150.9 & 2400+00 &      \\
 24.112+0.236 & 18 34 24.256 & ${-}$07 44 04.33 &      1.58 &       6.01 &    0.290 &   13.53 &    3.90 &   80.7 & 2400+00 &      \\
 24.199+0.243 & 18 34 32.417 & ${-}$07 39 14.22 &      1.74 &       8.63 &    0.331 &   17.31 &    7.85 &  157.4 & 2416+00 &      \\
 24.229+0.120 & 18 35 02.281 & ${-}$07 41 02.53 &      0.93 &       2.65 &    0.173 &   14.59 &    3.88 &   19.9 & 2433+00 &      \\
 24.363+0.044 & 18 35 33.554 & ${-}$07 35 58.98 &      1.33 &       9.47 &    0.247 &   18.72 &   11.52 &  159.4 & 2433+00 &      \\
 24.446${-}$0.168 & 18 36 28.354 & ${-}$07 37 27.76 &      0.94 &       2.43 &    0.185 &    8.95 &    5.47 &  142.2 & 2450${-}$33 &      \\
 24.456${-}$0.352 & 18 37 09.040 & ${-}$07 41 59.11 &      1.41 &       2.74 &    0.268 &    7.77 &    0.00 &   84.2 & 2450${-}$33 &      \\
 24.748${-}$0.206 & 18 37 10.158 & ${-}$07 22 23.62 &      2.22 &       2.56 &    0.413 &    4.92 &    0.00 &  137.0 & 2483${-}$33 & o    \\
 24.774+0.187 & 18 35 48.563 & ${-}$07 10 10.18 &      1.12 &       1.18 &    0.216 &    2.38 &    0.00 &  104.0 & 2483+00 &      \\
 24.814+0.122 & 18 36 06.960 & ${-}$07 09 50.03 &      2.18 &       3.04 &    0.435 &    4.42 &    0.00 &   80.0 & 2483+33 &      \\
 25.157+0.057 & 18 36 59.041 & ${-}$06 53 19.82 &      1.94 &      11.65 &    0.363 &   29.36 &    5.30 &   18.2 & 2533+00 &      \\
\enddata
\tablecomments{Table 6 is published in its entirety in the electronic edition of the {\it Astronomical
Journal} and is also available on the MAGPIS website (\url{http://third.ucllnl.org/gps}).  A portion is
shown here for guidance regarding its form and content.}
\tablenotetext{a}{
Source notes: `o' indicates that the source was
in the original catalog of Becker et al.\ (1994).
}
\end{deluxetable*}

\clearpage
\end{landscape}

\end{document}